# An interdisciplinary approach to high school curriculum development: Swarming Powered by Neuroscience

**Elise Buckley[1], Joseph D. Monaco[2], Kevin M. Schultz[1], Robert Chalmers[1], Armin Hadzic[1], Kechen Zhang[2], Grace M. Hwang[1,3*], M. Dwight Carr[1*]**

[1]Applied Physics Laboratory, Johns Hopkins University, Laurel, USA

[2]School of Medicine, Johns Hopkins University, Department of Biomedical Engineering, Baltimore, USA

[3]Kavli Neuroscience Discovery Institute, Johns Hopkins University, Baltimore, USA

**\* Correspondence:**

Grace M. Hwang                                          M. Dwight Carr
Grace.hwang@jhuapl.edu                          Dwight.carr@jhuapl.edu

**Keywords: high school, instructional technology, neuroscience, navigation, science education, technology, swarm robotics**

## Abstract

This article discusses how to create an interactive virtual training program at the intersection of neuroscience, robotics, and computer science for high school students. A four-day microseminar, titled Swarming Powered by Neuroscience (SPN), was conducted virtually through a combination of presentations and interactive computer game simulations, delivered by subject matter experts in neuroscience, mathematics, multi-agent swarm robotics, and education. The objective of this research was to determine if taking an interdisciplinary approach to high school education would enhance the students' learning experiences in fields such as neuroscience, robotics, or computer science. This study found an improvement in student engagement for neuroscience by 16.6%, while interest in robotics and computer science improved respectively by 2.7% and 1.8%. The curriculum materials, developed for the SPN microseminar, can be used by high school teachers to further evaluate interdisciplinary instructions across life and physical sciences and computer science.

## 1    Introduction

The development of multi-agent platforms with small-scale robotic vehicles is an exciting target of state-of-the-art autonomous systems engineering: many new applications may emerge from controlling large, distributed groups of inexpensive but agile vehicles. Unmanned robots are rapidly becoming a crucial technology for commercial, military, and scientific endeavors throughout the United States and across the globe. Critical future applications such as disaster relief and search & rescue will require intelligent spatial coordination among many robots spread over large geographical areas. However, several gaps exist in multi-agent robotic controllers: current communication and control frameworks need to be improved to provide the adaptiveness, resilience, and computational efficiency required for operating in complex and rapidly changing real-world conditions (Murray, 2007; Hamann et al., 2016; Chung et al., 2018; Yang et al., 2018). A team of researchers at the Johns

Hopkins University Applied Physics Lab (JHU/APL) and JHU/School of Medicine (SOM) explored whether neuroscience may offer insights to create a new class of multi-agent robotic controllers that could begin to address these aforementioned gaps. The team first secured a grant issued by the United States National Science Foundation (NSF) Integrative Strategies for Understanding Neural and Cognitive Systems Program (NCS) Program, titled "Spatial Intelligence for Swarms Based on Hippocampal Dynamics" (Whang, 2018). This project brought together principles from theoretical neuroscience (i.e., the analysis and investigation of theories about brain function) and multi-agent autonomous robotic coordination (swarms) to develop smarter controllers based on the concept of the cognitive map (see subheading on cognitive map and path integration).

The first premise of this project was that the world is constantly changing, and mammals have evolved the cognitive ability to plan new paths as needed while avoiding predators and seeking rewards. By contrast, autonomous robots are less robust, and often have difficulty operating in complex environments with changing conditions, such as in uneven terrain and moving obstacles. A second premise is that individual robots in a group are analogous to neurons in an animal's brain, which interact with one another to form dynamic patterns that collectively signal locations in space and time relative to brain rhythms (Hwang et al., 2021). The distribution of information across space and time has enabled a new paradigm of swarm control, in which swarms can automatically adapt to changes in the world in the same way that a mammal, in this case a rodent, can figure out which detour to take around an unexpected obstacle. Both premises resulted in research that explored the mathematical equivalence between fast time-scale learning in the brain and swarm motion in simulated robots in a framework developed by the research team, coined NeuroSwarms (Monaco et al., 2019, 2020). The NeuroSwarms controller was implemented in a well-established commercially available 3D virtual environment, Unity3D, by the research team.

There have been limited examples of STEM (science, technology, engineering, and mathematics) educators who collaborate with Ph.D. level researchers to create interdisciplinary high school curriculum and instructions in neuroscience, robotics, and computer sciences. This paper describes such an interdisciplinary approach between researchers who had these deep disciplinary expertises and seasoned educators from the JHU/APL STEM Program Management Office. Together, the team cooperatively developed an interdisciplinary four-day microseminar for Maryland high school students, titled Swarming Powered by Neuroscience (SPN), which was offered virtually in January 2021. One of the goals of the SPN microseminar was to determine if student interest in STEM would be enhanced, and if so, would interest be uniform across disciplines or targeted to a specific discipline. Student assessment surveys were administered before and after the SPN microseminar.

## 2    Pedagogical framework

The SPN microseminar was developed to enhance and expand on concepts in the physical and life sciences ordinarily taught in high school in the United States. To do this, STEM educators applied concepts from the United States Next Generation Science Standards (NGSS) (National Research Council, 2012)) which comprises core and component ideas in the life and physical sciences. The Core idea that is most closely aligned to the team's Spatial Intelligence project is LS1.D Information Processing, which describes what students should learn regarding how organisms detect, process, and use information to understand and navigate within their environment. These concepts naturally tie into the formation of a cognitive map and path integration (refer to 2.1 for more details). Furthermore, NGSS LS1.D explains that complex organisms convert information that is sensed from the environment into neural signals that control motor movement and decision making. Students should understand how external and internal stimuli enable an organism to interact with and understand its environment as a function of electromagnetic, mechanical, and chemical signals. More specifically, these interactions with external and internal stimuli can be explained in the context of how an animal is able to achieve path integration within the cognitive map. At the conclusion of lessons that are designed to fulfill the requirements of the NGSS, high school students are expected to understand that the brains of complex animals are divided into distinct regions and neural pathways that enable visual and auditory perception, guide motor movement, interprets perceptual information, and decision making.

The guiding principles behind the SPN microseminar are as follows: (1) Develop developmentally appropriate curricular materials and 3D tasks for high school students; (2) Develop engaging activities; (3) Inform students about careers in neuroscience and engineering; (4) Change students' attitudes toward STEM; (5) Provide access to cutting-edge STEM research for high school students from a variety of socioeconomic backgrounds; (6) Facilitate discussions between students and subject matter experts in neuroscience and engineering.

## 2.1 Cognitive Map and Path Integration

A sense of direction refers to an individual's ability to know, without explicit guidance, the direction in which they are or should be moving. This sense of direction is necessary in the creation of a cognitive map. The cognitive map is defined in terms of space relative to the external world, i.e., an allocentric reference frame or top-down view of the world. The cognitive map is about allocentric relationships between external objects and oneself. It can be characterized as a neural representation of the external spatial world that represents the distances and direction between places. It allows one to orient oneself within an environment, imagine oneself in different locations of the environment, and construct sequences embedded as paths within that environment. Every location in it produces a prediction to items or landmarks in the environment. This predictive ability is what allows one to do action selection in an environment, e.g., take a right turn after a landmark. To learn a cognitive map, an individual needs to be able to convert a reference frame relative to one's body (i.e., an egocentric reference frame) to an allocentric reference frame. The conversion of reference frames from egocentric to allocentric is essential in navigating the world and in creating detours to get around unexpected obstacles.

A family of specialized spatial neurons in the mammalian brain work together to create a cognitive map. A few of these neurons are summarized in Figure 1A. These spatial neurons are characterized by the rate at which neuronal action potentials are released in certain locations within an environment, or more typically referred to as the neuron's location-dependent firing rate. A place field is defined as the area where neurons fire the most in an environment (refer to the red patches in Figure 1A). Different types of spatial neurons perform different functions; these spatial neurons are located in a deep region of the brain, referred to as the hippocampal formation. Figure 1A uses a 2D environment to illustrate these spatial neurons: place cells typically exhibit elevated firing rate in one restricted region of an environment; border cells exhibit elevated firing rate at environmental boundaries such as vertical surfaces (e.g., cliff or a wall); grid cells exhibit elevated firing rate whenever an animal is located at one of the vertices of a periodic triangular array spread over a region; head direction cells exhibit elevated firing rate whenever an animals' head faces a particular direction relative to the environment; speed cells has a firing rate that correlates with the running speed of the animal. How these spatial neurons achieve sophisticated navigational strategies, such as path integration, is a hot topic of neuroscience research.

As shown in Figure 1B, path integration is an egocentric (self-centered) process by which an animal sums the vectors of distance and direction traveled from a start point (e.g., nest) to estimate its current position. Accurate path integration allows an animal to find a direct path back to the start location (Figure 1B, path shown in red dashes) versus retracing its steps to return to the start location (Figure 1B, path shown if one were to travel by way of the blue arrow). Path integration is an important evolutionary survival function. Animals from ants (Wittlinger et al., 2006) to rodents (Lisman et al., 2017), among other animals, are known to leave their nest to respectively search for food or for their young and immediately return to their nest by taking the shortest path owing to their ability to navigate by path integration.

A problem with path integration as a navigation strategy is that errors may accumulate, and if a large amount of error is accumulated, this may cause the animal to miss its target. Researchers can determine if animals path integrate by experimentally introducing errors to perturb their ability to path integration accurately. In Wittlinger's 2006 study, the researchers both reduced and extended the animals' leg lengths to demonstrate that animals would miss their nest proportionally to the amount of leg-length manipulation. More specifically, ants with shortened legs would stop to search for their nest approximately five meters before reaching the actual nest location, while ants with extended legs would travel past their nest by approximately five meters. Path integration operates on self-motion signals which may include (1) proprioception, which is based on information from muscles and joints about limb position; (2) motor efference copy, which is based on information from the motor system that tells the rest of the brain what movements were commanded; (3) vestibular inputs, which is based on information from the vestibular system. For example, the inner ear informs the rest of the brain about motion. Subsequently, deeper parts of the brain – thalamus – integrate all these streams of information to construct a high-level head-direction signal; (4) optic flow, which is based on changes in the visual scene that are projected onto the retina which are transformed into motion vectors of the external world around the head. Self-motion is integrated over time, but so are errors: thus, path integration must be corrected, or reset, to an absolute, world-centered (allocentric) spatial frame of reference using external cues. These external cues may include (1) visual landmark; (2) odor gradient, which is based on changes in olfactory intensity that indicates the animal's motion relative to a food source; (3) interaural time difference, which is based on differential arrival times of acoustic waves to the two ears which allow the angle (horizontal/azimuthal) to a sound source to be directly calculated by the auditory cortex. For additional information about path integration, interested readers are referred to this recent comprehensive review (Savelli and Knierim, 2019).

## 3    Learning Environment

### 3.1    Participants

The JHU/APL STEM Program Management Office Program Management Office (PMO) has hosted out-of-school time K–12 STEM programs for over 40 years. The STEM PMO consists of six full-time staff with degrees in education, math, science, and engineering. All STEM PMO staff have extensive experience making APL research accessible to elementary, middle, and high school students. APL's core programs include Girl Power, Maryland MESA (Mathematics, Engineering, and Science Achievement), STEM Academy, and ASPIRE, a high school internship program. These programs attract students from the Baltimore and Washington DC Metropolitan areas in the United States.

APL's STEM Academy is a series of afterschool project-based courses that 8th– 12th grade students take to learn about topics ranging from critical thinking to circuit design. Each course is developed in-house by the STEM Academy Specialist and an APL subject matter expert, then taught to students by an APL STEM volunteer working in that field. The STEM Academy enrolls over 700 students per year (84% minority and 53% female). STEM Academy was evaluated by Johns Hopkins University's Center for Research and Reform in Education in 2018 and found to be effective at increasing interest in STEM careers and the desire to choose a STEM major in college. Students who had completed at least one computer programming course through the STEM Academy and were in 9th–12th grade were recruited for the SPN microseminar. Thirty-five students participated in the SPN microseminar; they were surveyed anonymously before and after the microseminar. The demographics of student participants are captured in Supplementary Material Table 1.

## 3.2  Curriculum

The team applied a cross-cutting approach to NGSS LS1.D, resulting in the development of 8 hours of material – 4 hours of presentations and 4 hours of experimentation using the 3D simulation environment Unity3D. These materials were distributed over four 2-hour virtual lessons, which were held daily after school on 11–14 January 2021. Each lesson was structured with a lecture followed by an interactive programming experience. The interactive programming experience was designed to incrementally introduce the students to simulated robotic swarming controllers with increased functionality as the lessons progressed.

Lesson 1 was designed to ensure uniform coverage in neuroscience by all students. The first part of Lesson 1, "Neuroscience Basics: the neuron and nervous system", provided an overview of the overarching project followed by a general overview of neuroscience. The second part of Lesson 1, "Introduction to Mathematical Swarming", provided an overview of the mathematics that give rise to collective swarming behaviors in nature.

Lesson 2, "How do animal finds their way around living in the wild?", focused on the concept of spatial navigation including the spatial neurons involved in the formation of a cognitive map. Lesson 2 also explained the methods by which animals transform a sense of motion from external cues (e.g., landmarks, odor gradients) into a sense of location by summing the vector of distances and direction traveled from a start position, a process referred to as path integration or dead reckoning.

Lesson 3, "Cognitive Swarming: from oscillations, attractors, to collective spatial behaviors", focused on a comprehensive review of neuroscience, the hippocampus, and spatial navigation. Lesson 3 also explained the research team's brain-inspired swarming controller, NeuroSwarms (Monaco et al., 2020).

Lesson 4, "The Evolution of Neuroswarms", explained the concept of basic research and how that differs from translational research, reviewed prior lecture materials, introduced a Pytorch implementation of the NeuroSwarms controller, and discussed career planning for high school students, followed by an extended question and answer period.

All lecture materials including powerpoint files and recorded lectures are available upon request.

### 3.2.1 Interactive Programming Experience

Students are challenged to discover all rewards in the simulated environments as quickly as possible by adjusting the parameters presented in the SPN User Interface (UI) during gameplay. See

Supplementary Material Table 2 for a complete list of parameter definitions. The left panel of figure 2 contains the SPN UI which is composed of a Main Controller outlined in red (top left), an Allocentric View outlined in blue that presents the simulated environment from above (top right), and an egocentric view, outlined in black, that presents the perspective of each robot (bottom panel). The Main Controller houses large buttons to start or pause the game, as well as to reset parameter values. The SPN UI was designed to allow students to operate a complicated set of equations by using slider bars corresponding to controller parameters to gain an intuition for how each parameter would affect the overall simulated swarming behavior and elapsed time to reward discovery. At any time during gameplay, students can restart the game or reset all parameter values respectively via the RESTART or RESET VALUES button. Students can view the environment and associated sensing range (i.e., communication between robots which appears as a red line between any two robots) from the perspective of each robot by hitting the PREV BOT or NEXT BOT buttons. The bottom row of the Main Controller, labeled "REWARDS FOUND" tallies the number of rewards that are discovered. At the start of each game, eight white boxes are shown; when the reward corresponding to each box is discovered, the white box will turn red.

The Allocentric View within the SPN UI (box in upper-right corner outlined in blue) conveys important information about the position of each robot (shown in blue) and rewards including the history of the robotic swarm's search trajectory. Each reward is initially shown in white and upon discovery, the reward turns red. Each robot's memory of the search space is created by the color of the pixels within this Allocentric View. Black pixels represent areas that the robots have not searched, while different shades of gray represent searched areas. Lighter shades of gray represent areas recently searched, while darker shades of gray represent areas searched long ago. Students can develop swarming strategies based on the search patterns captured in the Allocentric View. Each robot's memory-based knowledge of the environment is uniquely dependent upon its search trajectory. Thus the Allocentric View presents an allocentric map of egocentric spatial memory for each robot. Users can access each robot's memory by hitting the PREV BOT or NEXT BOT buttons.

In this article, representative simulations are illustrated for each lesson (see Figures 2–5), which will be described in the remainder of this section. In the first interactive programming experience, eight robots were presented in the simulated world along with eight stationary rewards dispersed within the environment. In Lesson 1, the rewards were initialized as white circles that turn red only upon discovery. Once rewards are discovered by the swarming robots, the reward will remain stationary while each robot will continue to explore the rest of the environment.

Programming experience in Lesson 1 introduced the students to robotic swarms operated by conventional non-biologically inspired controllers, i.e., dynamic co-fields (Spears et al., 2004; Chalmers, 2005)**,** seeking stationary rewards. As shown in Figure 2, at t=0 seconds, robots are still grouped together (see blue squares in Allocentric View, upper-right corner of left panel) and zero rewards have been discovered. At t=65.851 seconds, robots have found one reward (see red square on Main controller in upper-left panel and red circle shown in the Allocentric View in upper-right). At t=221.067 seconds, all eight rewards have been found as shown by both the Main Controller and Allocentric View. The majority of the search area was searched, as exemplified by the different shades of grey in the Allocentric View (upper-right). To access this lesson, go to https://johnsam2.github.io/DayOne/index.html.

Programming experience in Lesson 2 introduced the students to conventional swarms operated by dynamic co-fields with mobile rewards that moved randomly. This modification required students to figure out which parameter had to be altered to account for mobile rewards versus the stationary rewards in Lesson 1. More specifically, when adjusting the "Search Decay Exponent" parameter, if the rewards being searched are stationary, it would not make sense to revisit previously searched locations. However, if the rewards are mobile, then there is a clear benefit to adjust the "Search Decay Exponent" variable to ensure that robotic agents revisit old locations over time. As shown in Figure 3, at t=26.984 seconds, one reward was found. At t=58.325 seconds, three rewards were found. At t=310.575 seconds, all eight rewards were found. Notice the differences in reward locations in the Allocentric View (upper-right) across these time points. Discovered rewards (red circles) remain stationary, while undiscovered rewards were mobile until found. To access this lesson, go to https://johnsam2.github.io/DayTwo/index.html**.**

Programming experience in Lesson 3 introduced robotic swarms that were driven by the NeuroSwarms controller instead of the conventional dynamic co-field method used in Lessons 1 & 2. Importantly, the NeuroSwarms controller features a circularly periodic phase variable that takes values from 0 to $2\pi$, corresponding to the angles that span a complete revolution of a circle or, equivalently, a complete cycle (or wave) of an oscillation in time. Thus, each agent maintains an internal phase state, driven by input strength, as a basis for interactive coupling with other nearby agents. This phase-based coupling elicits the sort of spontaneous synchronization, both in-phase (i.e., for interagent attraction) and anti-phase (i.e., for interagent repulsion), observed by Huygens for pendulum clocks anchored to the same wooden board. In our simulations, the phase of each robotic agent was represented by the robot's color, which was updated based on a periodic HSV colormap at every timestep. Additionally, the NeuroSwarms controller featured two new parameters, Swarming Weight and Reward Weight, which determined the spatial reach of, respectively, swarming and reward-approach behaviors. Specifically, Swarming Weight set the size of the local neighborhood within which agents interacted with each other, via phase-coupled attraction and repulsion as described above. Similarly, Reward Weight set the strength of an interaction that guided agents toward visible rewards in the environment, i.e., within an agent's unobstructed line-of-sight. As shown in Figure 4, the Allocentric View (upper-right) shows the location of each reward and the position of every robot (squares of different colors). At 20.451 seconds, no rewards had been discovered. At 47.843 seconds, four rewards had been discovered. By 55.885 seconds, all rewards had been found. The NeuroSwarms controller resulted in faster discovery of all eight rewards using default settings compared to the dynamic co-field controller from Lessons 1 and 2. To access this lesson, go to https://johnsam2.github.io/DayThree/index.html.

Programming experience in Lesson 4 introduced the NeuroSwarms controller with practical parameters such as speed, sensor, and communication range (see Figure 5, orange portion between the Main Controller and the Allocentric View). All other control parameters were implemented identically as in Lesson 3. This progression allowed students to experience the intuitive difference between controllers inspired by neuroscience versus those that were not at an increasing level of engineering fidelity. As shown in Figure 5, at t=0 second, robots are in starting formation. At t=23.636 seconds, two rewards were found. At t=134.029 seconds, all eight rewards were found. To access this lesson, go to https://johnsam2.github.io/DayFour/index.html.

### 3.2.2 Interactive Programming Experience – Technology Challenges

The SPN microseminar was originally planned to take place in-person with the students working in small groups on live programming challenges related to dynamic co-fields and NeuroSwarms on laptops with preloaded programs. The network would be mapped together in "land parties" so that each small team could explore one region of the island. A culminating activity would include all the teams exploring a new portion of the island simultaneously and seeing which team could maximize their rewards through applying the concepts they learned during the SPN microseminar.

The SPN microseminar was originally conceived to be a 200 student activity to be held in-person on one weekend in August of 2020. However, due to the worsening conditions of the COVID-19 pandemic, it became necessary to convert this in-person event to a virtual forum which was held January 11-14, 2021, over two-hour increments in the late afternoon. The decision to transform to a virtual environment posed technical challenges to the software development team and the educational team. The educational priority was that the students would still be able to work in a hands-on environment where they would be able to manipulate the parameters and directly see the effect on the swarming robots. In planning this virtual version of the microseminar, it was essential that any student who was eligible would be able to participate in the microseminar regardless of the type of computing device they had in their home. Whereas originally the students would be working in-person on laptops with the programs pre-loaded and tested. As a consequence, the software team needed to develop an environment that could be downloaded and provide troubleshooting remotely. Most STEM Academy students attend public schools in Maryland, where many districts provided Chromebooks, or equivalent computing devices, to their students for virtual learning.

To ensure equity of access to the microseminar, the software development team's new goal was to develop an environment that would be functional within the computing limitations of a Chromebook. This meant that access to the software developers for in-person live troubleshooting was no longer viable. Instead, the application needed to be distributable on a machine made to stream apps in chrome rather than anything that required too much processing power. There would only be limited virtual technical support for the students, so the swarming environment for the students to access needed to be as streamlined as possible.

The software development team focused on coding in WebGL, which allows for interactive graphics on web browsers, such as Chrome. Not having the time or resources to build a new website, which would need to be maintained for longevity, the software team evaluated hosting options and settled on putting the environments live in GitHub. The software team developed four interactive environments where the students were able to individually control the swarms. Environments can be found at the following links: https://johnsam2.github.io/DayOne/index.html, https://johnsam2.github.io/DayTwo/index.html, https://johnsam2.github.io/DayThree/index.html, https://johnsam2.github.io/DayFour/index.html. In planning the microseminar via Zoom, time was set aside on the first day for any technology troubleshooting and a Zoom breakout room was staffed by a member of the software development team to address any technical concerns.

### 3.2.3 Evaluations

The students were surveyed about their interest in various STEM careers prior to and at the conclusion of the microseminar. They were asked to indicate their interest in pursuing careers in Neuroscience, Robotics, and Computer Science by selecting a number on a sliding scale from 1-100. Percent change in interest in pursuing a career in STEM was computed based on the difference between the average "Percent interest: Pre microseminar" and the average "Percent interest: Post microseminar" normalized by the average "Percent interest: Pre microseminar" for each discipline (see Supplementary Material Table 3). Thirty-five students responded to the pre microseminar survey, while seventeen students responded to the post microseminar survey. Because student identities were protected, it was not possible to compute paired statistics.

### 4   Results

### 4.1   Student Survey Summary

Quantitative results of student interest in STEM topics are summarized in Supplementary Material Table 3 on the basis of pre- and post-SPN microseminar participation. The most notable percent increase in interest was in the field of neuroscience, which had students indicating a 16.6% change in their interest. Modest increases in interest were observed in the fields of robotics and computer science corresponding respectively to 2.7% and 1.9%. This is an encouraging result because many students had little to no familiarity with neuroscience prior to the SPN microseminar. It is also notable that these students were recruited from the STEM Academy, which meant each student had already successfully completed at least one course in a computer programming language and had prior exposure to robotics. Thus it was not surprising that their relative interests in these fields were smaller than in neuroscience. Notably, the SPN microseminar did not negatively affect their interest in robotics and computer science, and instead substantially elevated their interest in neuroscience.

Qualitative results of student comprehension regarding the concept of NeuroSwarms are summarized in Supplementary Material Table 4 on the basis of pre- and post-microseminar participation. Prior to the microseminar, students were asked to define the concept NeuroSwarms without any internet help. Many interesting responses were provided, a sample of them are presented in Supplementary Material Table 4. Out of the thirty-five responses, two students admitted that they did not know the answer (not shown in Supplementary Material Table 4), while one student provided a correct answer (bold text in Supplementary Material Table 4, first column). A few students imagined that NeuroSwarms was a brain-computer interface controller, which was a reasonable but incorrect guess. After the microseminar, thirteen out of seventeen survey respondents provided a range of correct answers. This demonstrates that 76% of the survey respondents understood the lecture on NeuroSwarms.

Quantitative results of student satisfaction by STEM topics and interdisciplinary teaching experiences are summarized in Supplementary Material Table 5 based on post-microseminar participation. The percentage of students who strongly agreed that they learned new knowledge in disciplinary topics varied – i.e., neuroscience: 82%, robotics: 53%, and computer science: 41%. The majority of students (64%) strongly agreed that they enjoyed learning from an interdisciplinary team of experts and 70% strongly agreed that the microseminar emphasized the need to have instruction teams with diverse disciplinary backgrounds.

## 4.2 Interactive Programming Experience

During the actual SPN microseminar, every student was able to access the interactive Unity3D environment virtually. In fact, no student visited the microseminar's technology support breakout room. The interactive environment worked despite all the students accessing at the same time, allowing for the students to share their insights and top scores via chat in Zoom, thus retaining an element of the original competition-based approach to the NeuroSwarms Unity3D environment.

## 5 Discussion

The SPN microseminar illustrates an effective way to teach neuroscience to high school students who already have a background in robotics and computer science. Our results suggest that the materials developed for the microseminar and the microseminar format could be used as an effective model for developing lesson plans and a strategy to involve STEM professionals in the delivery of instruction to attain the learning objectives of LS1.D. Future microseminars could explore methods to engage robotics and computer science to students who already have a background in neuroscience to determine if reciprocity in interdisciplinary learning experiences is observed.

## 6    Conflict of Interest

*The authors declare that the research was conducted in the absence of any commercial or financial relationships that could be construed as a potential conflict of interest.*

## 7    Author Contributions

EB, JM, KS, RC, KZ, GH, MC conceived the study. RC developed the simulations presented in this study. LB, JM, KS, AH, GH designed and presented the lectures. EB performed the analysis of student surveys. EB, GH, and MC wrote the manuscript. All authors read and approved the content.

## 8    Funding

This work was supported by National Science Foundation Award NCS/FO 1835279, and the Johns Hopkins University Kavli Neuroscience Discovery Institute.

## 9    Acknowledgments

This material is based on work supported by (while serving at) the National Science Foundation. Any opinion, findings, and conclusions or recommendations expressed in this material are those of the authors and do not necessarily reflect the views of the National Science Foundation.

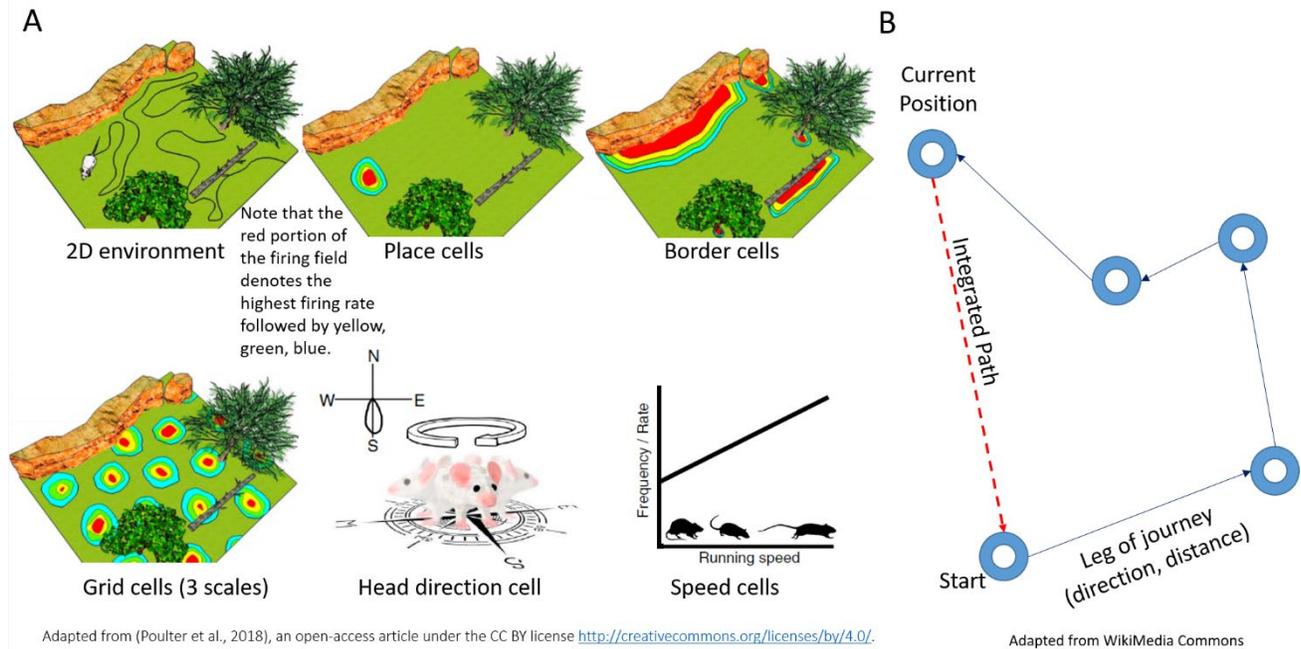

Figure 1. A. Spatial neurons that constitute the cognitive map. Adapted from (Poulter et al., 2018), an open access article under the CC BY license http://creativecommons.org/licenses/by/4.0/. B. Path integration is the process by which an animal sums the vectors of distance and direction traveled from a start point (e.g., nest) to estimate its current position. Accurate path integration allows an animal to find a direct path back to the start location (red dashes) versus retracing one's steps to return to the start location (blue arrows). Adapted from WikiMedia Commons.

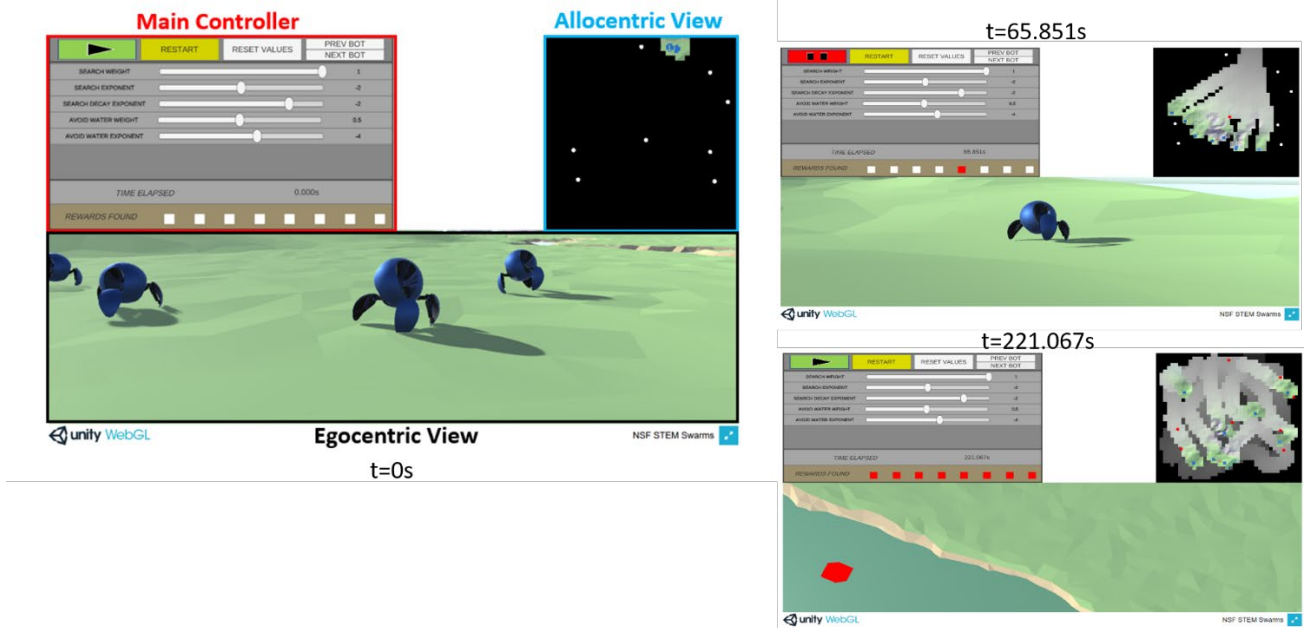

Figure 2. Programming experience in Lesson 1 – Robots controlled by Dynamic Co-field controller with stationary rewards. Swarming Powered by Neuroscience – User Interface illustrated in left panel.

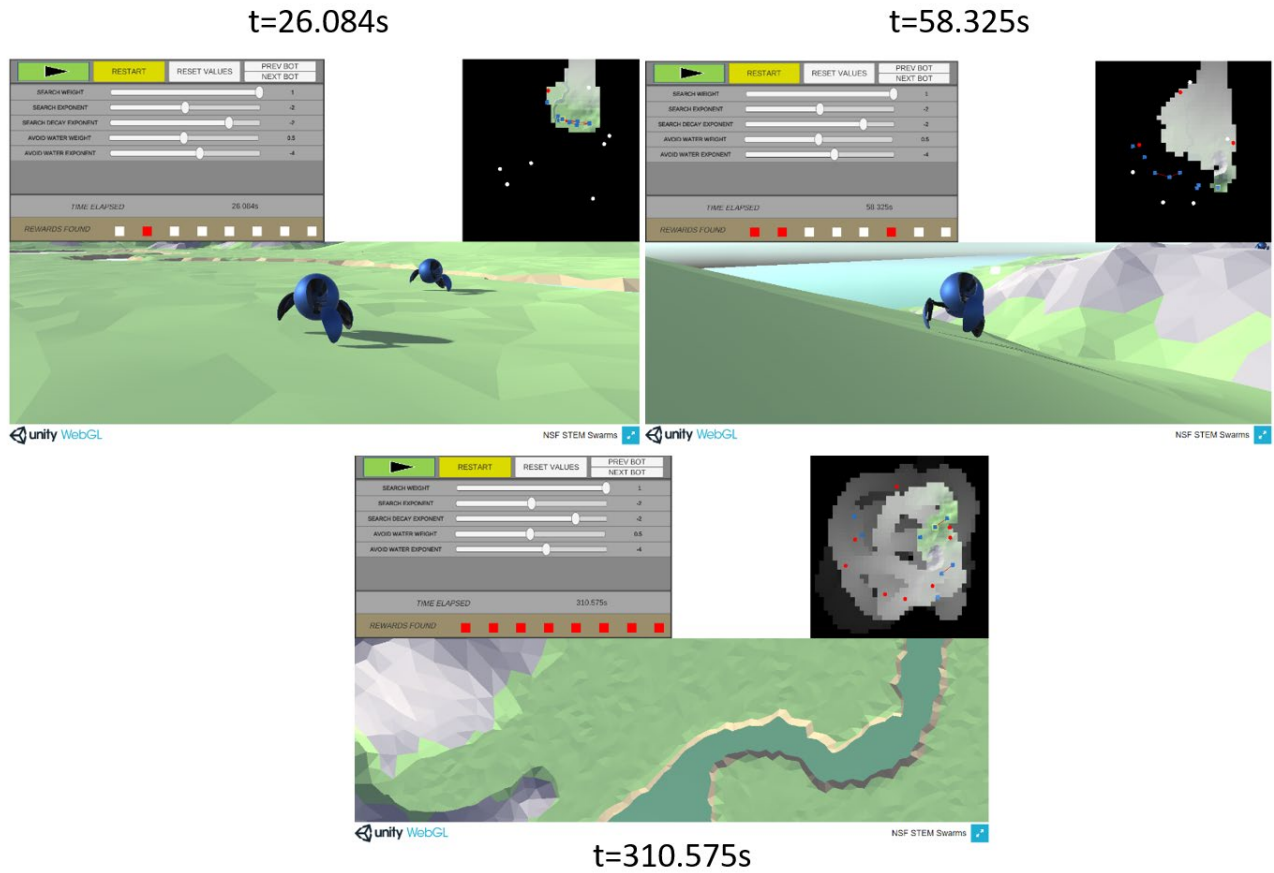

Figure 3. Programming experience in Lesson 2 – Robots controlled by Dynamic Co-field controller with mobile rewards.

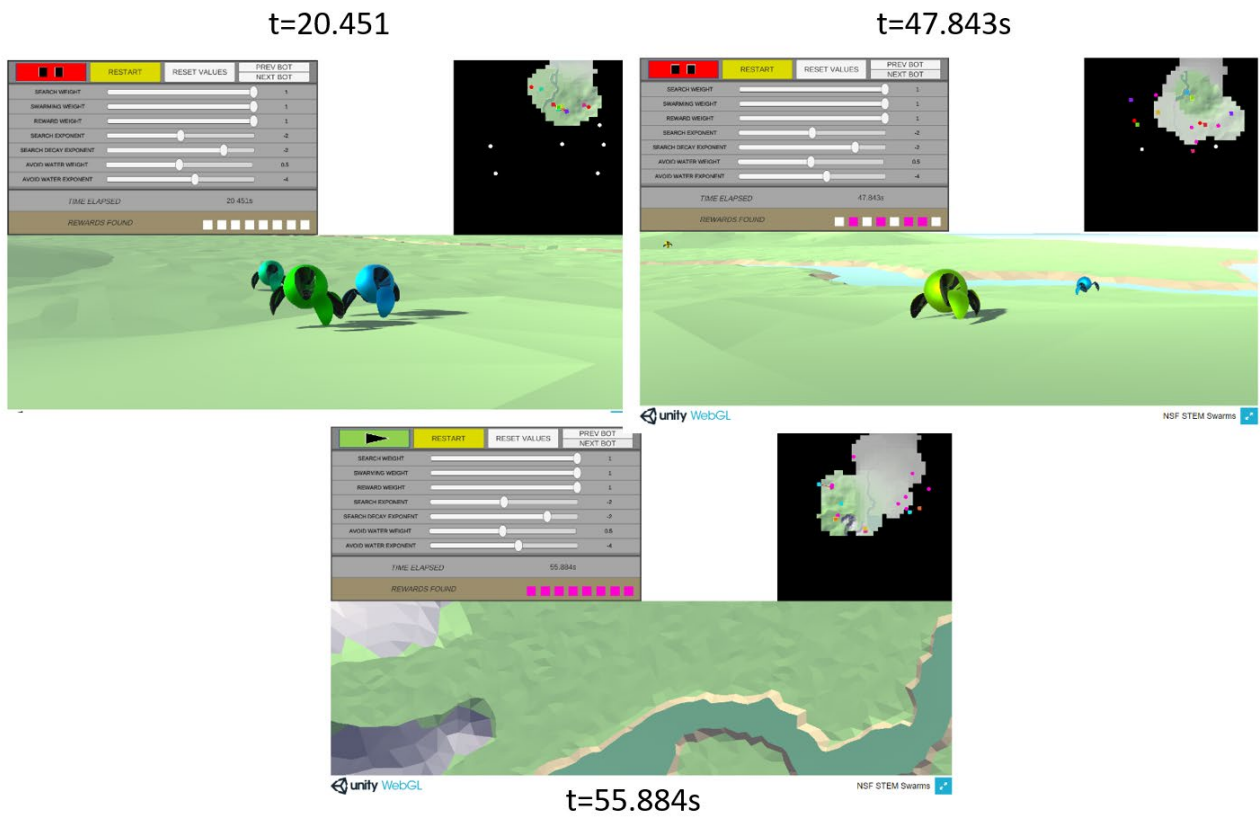

Figure 4. Programming experience in Lesson 3 – Robots controlled by the NeuroSwarms controller with mobile rewards.

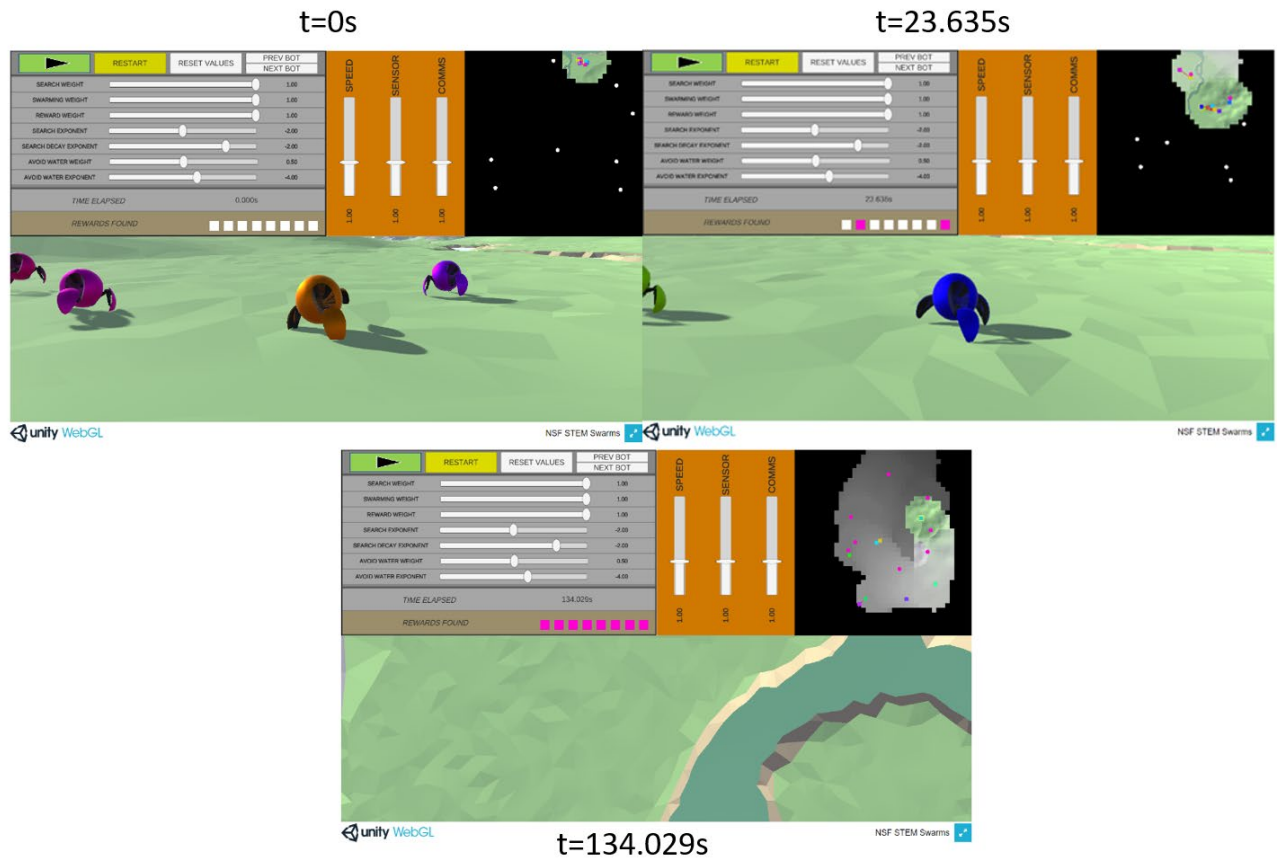

Figure 5. Programming experience in Lesson 3 – Robots controlled by the NeuroSwarms Controller with mobile rewards and adjustable speed, sensors, and communication.

# *Supplementary Material*

Table 1. Swarming Powered by Neuroscience Microseminar Student Demographics.

| Demographic Category | Percent of Participants |
|---|---|
| Female | 31.4% |
| Male | 65.7% |
| Not disclosed | 2.9% |
| Asian | 37.1% |
| Black or African American | 11.4% |
| Hispanic or Latino | 2.9% |
| White | 22.9% |
| Two or more races | 8.6% |
| Not disclosed | 17.1% |
| 9th grade | 42.9% |
| 10th grade | 25.7% |
| 11th grade | 14.3% |
| 12th grade | 17.1% |

Table 2. Simulated Robotic Controller Parameters.

| Controller Parameter [range] | Description | Default |
|---|---|---|
| Search Weight [0–1] | The value by which vectors in the controller are weighted by prior to vector summation to affect the overall behaviour of each robot. | 1 |
| Search Exponent [-5–1] | Logarithmic control parameter of the physical range of the robots. Lower values make it more likely that robots will spread out over the environment. | -2 |
| Search Decay Exponent [-10–0] | Refers to how quickly a robot forgets what it has seen previously. | -2 |
| Avoid Water Weight [0–1] | Same as Search Weight when a robot enters the ocean. | 0.5 |
| Avoid Water Exponent [-10–0] | Same as Search Exponent when a robot enters the ocean. | -4 |
| Swarming Weight [0–1] | Spatial constant that sets the size of the interagent interaction kernel in NeuroSwarms simulations. Lower/higher values increase/decrease local clustering and decrease/increase global exploration of the environment. | 1 |
| Reward Weight [0–1] | Spatial constant that sets the attraction strength toward visible rewards in NeuroSwarms simulations. Lower/higher values decrease/increase the range of distances from which agents can approach visible rewards. | 1 |
| Speed [0–3] | The speed of each robot. | 1 |
| Sensor [0–3] | Each robot's sensor range. | 1 |
| Comms [0–3] | Each robot's communication range. | 1 |

Table 3. Student Survey Results.

| Career Field | Percent Interest: Pre microseminar | Percent Interest: Post microseminar | Percent Change in Interest in Pursuing Career in STEM |
|---|---|---|---|
| **Neuroscience** | 51.1% | 59.6% | + 16.6% |
| **Robotics** | 66.2% | 68% | + 2.7% |
| **Computer Science** | 79.9% | 81.4% | + 1.9% |

Table 4. Describe what you think a "neuroswarm" is. (Do not google).

| Pre microseminar | Post microseminar |
|---|---|
| • Controlling a swarm or group of things or robots with the brain.<br>• Overwhelming information going into the brain or control center.<br>• What I think that "neuroswarm" means is using tiny robots and computer to find and solve neurological problems.<br>• A swarm of neuro diseases happening at the same time in your brain.<br>• A large group of micro processors that act and solve problems like a real brain.<br>• A grouping of neurons in the brain the work as part of a swarm to achieve goals.<br>• I think a "neuroswarm" might be what you call the field of robotics related to neuroscience.<br>• **A robotic swarm that works like a brain.**<br>• I think it is improving network or data intelligence by adapting to the bio mimicry mechanisms for real world problems.<br>• When many (hundreds of) robots that are programmed, will complete tasks and use machine learning to learn from experience in order to more efficiently complete the task. | • A group of robots that have depend on each other for actions through a neural similar system.<br>• Neuroswarm is the network of robots that is based on the concept of neuron connectivity in animals to study how decisions are made. It is very [interesting] bio mimicry concept<br>• A neuroswarm is a group of robots that communicate with one other to complete a task<br>• A swarm of individual robots that act like the hippocampus or hippocampus-related structures in the brain.<br>• It is a swarm that has a system modeled after how neurons work specifically in the hippocampus.<br>• I think that a neuroswarm is a group of entities that "think" and act without the input of a human to accomplish a preset goal. They form swarming patterns to accomplish their goal and if implemented in robots, the "neurons" in the swarm can communicate with one another.<br>• A neuroswarm is when multiple things, such as robots, or virtually coded things do things together to help the whole. |

Table 5. Student Satisfaction Survey Results Post microseminar.

| Question | Disagree | Neither agree or disagree | Agree | Strongly Agree |
|---|---|---|---|---|
| • I learned about an application of **neuroscience** I had not heard of before | 0 | 0 | 3 | 14 |
| • I learned about an application of **robotics** I had not heard of before | 1 | 0 | 7 | 9 |
| • I learned about an application of **computer programming** I had not heard of before | 0 | 3 | 7 | 7 |
| • I enjoyed learning from an **interdisciplinary** team of experts | 0 | 0 | 6 | 11 |
| • This microseminar emphasized the need to have teams with **diverse** disciplinary backgrounds | 0 | 1 | 4 | 12 |